\documentclass[twocolumn,showpacs,preprintnumbers,amsmath,amssymb]{revtex4-1}
\usepackage{graphicx}
\usepackage{dcolumn}
\usepackage{bm}

\begin{document}

\newcommand{\be}{\begin{eqnarray}}
\newcommand{\ee}{\end{eqnarray}}
\newcommand{\nn}{\nonumber\\}
\newcommand{\nin}{\noindent}
\newcommand{\la}{\langle}
\newcommand{\ra}{\rangle}

\renewcommand{\theequation}{\arabic{section}.\arabic{equation}}

\title{Antiferromagnetic, metal-insulator, and superconducting
phase transitions in underdoped cuprates: 
Slave-fermion  $t$-$J$ model in the hopping expansion}

\author{Akihiro Shimizu$^\ast$} 
\author{Koji Aoki$^\ast$} 
\author{Kazuhiko Sakakibara$^\star$}
\author{Ikuo Ichinose$^\ast$}
\author{Tetsuo Matsui$^\dag$}
 \affiliation{${}^\ast$Department of Applied Physics, Graduate School of 
Engineering, \\
Nagoya Institute of Technology, 
Nagoya, 466-8555 Japan 
}
\affiliation{%
${}^\star$Department of Physics, Nara National College of Technology, 
Yamatokohriyama, 639-1080 Japan
}%
\affiliation{%
${}^\dag$Department of Physics, Kinki University, 
Higashi-Osaka, 577-8502 Japan
}%

\date{\today}

\begin{abstract}
In the present paper, we study a system of doped antiferromagnet
in three dimensions at finite temperatures
by using the $t$-$J$ model, a canonical model of strongly-correlated 
electrons.
We employ the slave-fermion representation of electrons in which
an electron  is described as a composite of a charged spinless holon
and a chargeless spinon. 
We introduce two kinds of U(1) gauge fields on links as auxiliary fields, 
one describing resonating valence 
bonds of antiferromagnetic nearest-neighbor spin pairs and
the other for nearest-neighbor hopping amplitudes of
holons and spinons in the ferromagnetic channel.
In order to perform numerical study of the system, we integrate out
the fermionic holon field by using the hopping expansion
in powers of the hopping amplitude, 
which is legitimate for the region in and near the insulating phase. 
The resultant effective model is described in terms of bosonic
spinons and the two U(1) gauge fields, and a collective field
for hole pairs.
We study this model by means of Monte-Carlo simulations,
calculating  the specific heat, spin
correlation functions, and  instanton densities.
We obtain a phase diagram in the hole concentration-temperature plane,
which is in good agreement with that observed
recently for clean and homogeneous underdoped samples.
\end{abstract}

\pacs{74.72.-h, 11.15.Ha, 74.25.Dw}

\maketitle

\section{Introduction}
\setcounter{equation}{0} 
Since the discovery of  high-temperature superconductors of cuprates,
it has passed more than two decades\cite{bednorz}. 
Besides their high critical temperatures ($T$) of superconducting (SC)
phase transition,
these cuprates have several interesting properties like anomalous
properties in the metallic state,  existence of Fermi arcs, etc\cite{exp}.
To explain these properties, various theoretical approaches have 
been proposed\cite{anderson}. 
Although ample knowledge have been accumulated,
we still do not have a theory that has been proved and accepted 
as a ``right" one.

The $t$-$J$ model\cite{tj} is regarded as 
one of the canonical models for  high-$T_c$ cuprates.
The model excludes doubly-occupied electron states at each site
reflecting the strong correlations (Coulomb repulsion) 
among electrons.  This condition makes it hard to get convincing
and solid  understanding of the model such as its phase structure. 

The slave-particle approach\cite{slave} using  
the slave-fermion (SF) or the slave-boson representation 
has been proposed in order to treat  this local constraint 
on the physical states faithfully.
In the SF representation, each  electron is described as a composite of 
a charged spinless fermionic particle called holon and a neutral 
bosonic particle with spin called spinon.
The SF approach is known to be superior\cite{yoshioka} (giving
a lower ground-state energy in the mean-field theory)
than its statistics-reversed assignment, the slave-boson representation, 
in the region with small hole concentrations $\delta$ ($\delta$ is
just the density of holons per site).

We have studied the SF $t$-$J$ model in path-integral formalism\cite{im}.
Let us summarize the results of Ref.\cite{im}.
The local constraint is exactly respected by using the CP$^1$ 
(complex projective) variables (we write it $z_{x\sigma}$ in Sect.II)
for spinons.
The fermionic holons ($\psi_x$) are described by Grassmann numbers.
In path-integral expression of the partition function, 
fluctuations of variables along the 
imaginary time give rise to certain 
imaginary term in the action.
By assuming the short-range antiferromagnetic (AF) order between the 
nearest-neighbor (NN) spin pair, we integrated over a half of 
the spinon variables, those sitting at the odd sites, 
assuming a short-range (SR) AF order to obtain an effective model.
At the half filling ($\delta=0$) the effective model reduces to
the Heisenberg spin model. 
It favors the so-called resonating
valence bonds (RVB), the NN spin singlet pairs with AF coupling.   
As holons are doped, 
the AF order are gradually destroyed because hopping of holons
is associated by hopping of spinons without spin flips, which
breaks some RVB's. 

Also there arises 
an attractive force between NN holon pair reflecting the energy released
by breaking RVB's. In fact, the NN holon pair breaks only 7(11) RVB's while
a holon pair separated at longer distance breaks 8(12) RVB's
in two(three)-dimensional lattice.
In Ref.\cite{im}, we have introduced a hole-pair field, 
the condensation of which implies the SC state,
and derived its Ginzburg-Landau (GL) model 
in the hopping expansion. 
At the mean-field level, this GL favors 
the so-called flux phase corresponding to $(s+id)$-wave symmetry.

The slave-particle approach intrinsically possesses
U(1) gauge symmetry, since the electron
operator is invariant under the local and simultaneous 
rotation of phases of holon and spinon fields.
The possible charge-spin separation phenomena\cite{css}
has a natural and potentially simple explanation
such that the U(1) gauge dynamics in cuprates is realized
in the deconfinement phase. In fact, in the deconfinement phase,
holons and spinons may appear as unbound quasiparticles
moving independently
due to the weak gauge force among them.

The slave-particle approach has yet another advantage.
The mean field theory based on the slave-particle representation
is basically capable to describe various expected phases including
the SC phase\cite{mft}.
However, the criticism to this result may be common to every mean 
field theory, i.e., the faithful evaluation of effects of fluctuations
around mean fields are missing. 
It is rather hard to evaluate such effects analytically
in nonperturbative manner 
because the model has local gauge symmetry as mentioned and associated
zero modes may give rise to strong effects in the infrared region. 

One may think that numerical studies may be one viable approach
as the successful example  of  lattice gauge theory in 
high-energy physics demonstrates.
However, straightforward numerical studies 
such as Monte-Carlo (MC)
simulations of the SF $t$-$J$ model 
in path-integral representation  are still not feasible
because of the notorious sign problem in the fermionic determinant 
generated upon integrating over holon variables.

In the present paper, we shall revisit the $t$-$J$ model
on a three-dimensional (3D) lattice
in the SF path-integral representation with the purpose 
to study its nonperturbative aspects by numerical methods. 
In order to avoid the difficulty associated with
fermionic determinant mentioned above, we derive
an effective model by employing the hopping 
expansion to evaluate integrals over fermionic holons.
It is an expansion
in powers of the hopping amplitude of holons.
An effective expansion parameter is  $t\times \delta$,
so the expansion is useful and legitimate at sufficiently 
low dopings $\delta$. 
In the region of applicability of the hopping expansion, the wild behavior
of fermionic determinant is suppressed in a natural way.
We respect the structure of interaction terms
generated by the hopping expansion, but  consider their coefficients 
as independent parameters in a flexible manner. This is partly because
these coefficients acquire renormalization via higher-order terms
in the expansion.    

As mentioned, the action in path-integral representation
involves the imaginary part reflecting the imaginary-time dependence 
of the variables. This brings some complications in
numerical approach.  In Ref.\cite{im}
we have seen that the integration over odd-site spinon variables
makes the spinon part of the resultant action real.
In this paper, we avoid this imaginary part in another manner
by simply considering the region of finite $T$'s;
at sufficiently high $T$ such that the dependence of variables 
on the imaginary time may be neglected.
We expect that each  phase obtained at finite $T$'s
survives down to  sufficiently low $T$'s including $T=0$,
so the obtained phase diagram at finite $T$'s is useful
not only for itself but also for low $T$'s down to $T=0$.

In the practical numerical study, knowledge and techniques 
developed in the study of lattice gauge theory of
high-energy physics are helpful. 
By making MC simulations of the effective model,
we obtain a phase diagram in the  
$\delta$-$T$ plane, which contains AF phase, 
SC phase, and metal-insulator (MI) transition. 
The overall phase structure
is similar to that observed in experiments for 
lightly-doped materials\cite{exp2}.

The present paper is organized as follows.
In Sect.II, we explain and set up the model in detail.
The holon variables are analytically integrated out 
by means of the hopping expansion to obtain
the effective model at small $\delta$'s and finite $T$'s. 
The model includes several variables; (i) the spinon field $z_{x\sigma}$, 
(ii) the auxiliary field for spin-singlet (RVB) amplitude
of NN spinon pair (we call it $U_{x\mu}$
in Sect.II),
(iii) the auxiliary field for amplitude
of holon and spinon hoppings in the ferromagnetic (FM) channel 
($V_{x\mu}$), which works as an order parameter of the MI transition, 
and (v) the hole-pair field ($M_{x\mu}$) for superconductivity.
We introduce the hole-pair field and include the associated
GL terms to the effective action, respecting the NN attractive force 
between holons as discussed in Ref.\cite{im}.

In Sect.III, we first study the case without the superconducting channel
(by neglecting the GL energy of hole-pair field).
We present the results of MC  simulations
for the corresponding model, which we call $UV$ model. 
We calculated spin correlation function to
study the AF transition, and 
instanton densities of $U$-field and $V$-field to study 
the decay of AF order and the MI transition.
We locate the AF and MI phase transition lines.

In Sect.IV, we study the full model including
the SC channel and discuss SC phase transition together with
AF and MI ones.
We modify the coefficients of GL terms of hole pairs
from the leading-order values of hopping expansion of Ref.\cite{im}
so as to describe the  $d$-wave SC observed in experiments 
instead of $s+id$ one. 
This is because the SC transition is expected (and actually verified later on)
to occur in the metallic phase and the higher-order terms
of the hopping expansion should be included.
Our standpoint is that we regard the hole-pair
part of the effective model in a flexible manner, i.e.,
its structure is suggested by hopping expansion but
its coefficients are relaxed to study
the region beyond the validity of the leading order of the
hopping expansion.
We find that the SC state occurs always in the metallic phase, whereas
the AF long-range order (LRO) can coexist with the SC.
There appear two phase transitions related with the SC.
One is a primordial SC transition that stabilizes the amplitude of 
hole pairs and gives rise to a pseudo-gap in holon excitation energy,
while the other is a genuine SC transition reflecting
a phase coherence of hole pairs associated with the Higgs mechanism.

In Sect.V we present  discussions and  conclusions.
We discuss that the present model offers us an interesting possibility of
new description of a SC state
in the framework of gauge theory with {\it local interactions}. 
 
In Appendix A we give some details of the hopping expansion of
path-integral over holons.   

\section{The $t$-$J$ model in the slave-fermion representation 
and holon hopping expansion}
\setcounter{equation}{0} 

\subsection{Path integral expression}
We  start with the standard $t$-$J$ model on 
a 3D cubic lattice\cite{3dasymmetry},
whose Hamiltonian is given  in terms of 
electron operator $C_{x\sigma}$ at site $x\ (=x_1,x_2,x_3)$
and spin $\sigma\ [\ =1(\uparrow), 2(\downarrow)]$ as follows;
\be
H&=&-t\sum_{x,\mu,\sigma}\big(\tilde{C}^\dagger_{x+\mu,\sigma}
\tilde{C}_{x\sigma}+{\rm H.c.}\big) \nn
&&+J\sum_{x,\mu}\Big[\vec{S}_{x+\mu}\cdot\vec{S}_x
-{1 \over 4}n_xn_{x+\mu}\Big],
\label{htj}
\ee
where
\be
\tilde{C}_{x\sigma} &\equiv& (1-C^\dagger_{x\bar{\sigma}}
C_{x\bar{\sigma}})\ C_{x\sigma},\nn
\vec{S}_x&\equiv&\frac{1}{2}\sum_{\sigma,\sigma'}C^\dagger_{x\sigma} 
\vec{\sigma}_{\sigma\sigma'}C_{x\sigma'},\ \ 
(\vec{\sigma}:{\rm Pauli\ matrices}), \nn
n_x&\equiv&\sum_{\sigma}C^\dagger_{x\sigma}C_{x\sigma}.
\ee
$\mu (=1,2,3)$ is the 3D direction index and also denotes
the unit vector. $\bar{\sigma}$
($\bar{1}\equiv 2, \bar{2}
 \equiv 1)$ denotes the opposite spin.
The doubly occupied states ($C^\dag_{x\uparrow}
C^\dag_{x\downarrow}|0\ra$) 
are excluded from the physical states due to the strong on-site Coulomb
repulsion. 
The operator $\tilde{C}_{x\sigma}$ respects this point.

We adopt the {\em slave-fermion representation}
of the electron operator $C_{x\sigma}$ as a composite form,
\be
C_{x\sigma} = \psi^\dagger_x a_{x\sigma},
\ee 
where
$\psi_x$ represents annihilation operator of the 
 fermionic holon carrying the charge $e$ and no spin and
$a_{x\sigma}$ represents annihilation operator of the bosonic 
spinon carrying $s=1/2$ spin and no charge.
Physical states $|{\rm Phys}\ra$ satisfy the following constraint, 
\begin{equation}
\Big(\sum_\sigma 
a_{x\sigma}^\dagger a_{x\sigma} +\psi_x^\dag \psi_x\Big)
|{\rm phys}\rangle
 = |{\rm phys}\rangle.
\label{const1}
\end{equation}
In the salve-fermion representation, the Hamiltonian (\ref{htj})
is given as 
\begin{eqnarray}
&&H=-t\sum_{x,\mu}\Big(\psi_x^\dagger a_{x+\mu}^\dagger a_x\psi_{x+\mu}
+\psi_{x+\mu}^\dagger a_{x}^\dagger a_{x+\mu}\psi_{x}
\Big)\nonumber \\
&&\quad\quad+{J \over 4}\sum_{x,\mu}\Big[(a^\dagger\vec{\sigma}a)_{x+\mu}\cdot
(a^\dagger\vec{\sigma}a)_x-(a^\dagger a)_{x+\mu}(a^\dagger a)_x\Big],
\nn
&& (a^\dagger a)_x\equiv\sum_\sigma 
a^\dagger_{x\sigma}a_{x\sigma},\
(a^\dagger\vec{\sigma} a)_x\equiv \sum_{\sigma,\sigma'} 
a^\dagger_{x\sigma}\vec{\sigma}_{\sigma\sigma'}a_{x\sigma'}.
\label{htj2}
\end{eqnarray}
We employ the path-integral expression
for the partition function of the $t$-$J$ model,
\be
Z = {\rm Tr}\exp(-\beta H),\quad \beta\equiv \frac{1}{k_{\rm B}T},
\ee
at finite $T$ in the slave-fermion representation.
This is done by introducing a complex number 
$a_{x\sigma}(\tau)$  and a Grassmann number $\psi_x(\tau)$
at each site $x$ and the imaginary time 
$\tau \in [0,\beta]$.
The constraint (\ref{const1}) is solved\cite{im} 
by introducing
CP$^1$ spinon variable $z_{x\sigma}(\tau)$,
i.e., two complex numbers $z_{x1}, z_{x2}$ for each site $x$
satisfying 
\be
\sum_\sigma \bar{z}_{x\sigma} z_{x\sigma} = 1,
\label{z}
\ee
and writing 
\be
a_{x\sigma} = (1-\bar{\psi}_x\psi_x)^{1/2}z_{x\sigma}.
\label{a}
\ee
It is easily verified that the constraint (\ref{const1})
is satisfied by Eqs.(\ref{a}) and (\ref{z}).
Then, the partition function 
in the path-integral representation is given by
an integral over the
CP$^1$ variables $z_{x\sigma}(\tau)$ and Grassmann numbers 
$\psi_x(\tau)$.

We shall consider the system at finite and relatively high
$T$'s, such that
{\it the $\tau-$dependence of the variables $z_{x\sigma}, 
\psi_x$ are negligible} (i.e., only their zero modes survive).
Then the kinetic terms of $z_{x\sigma}, \psi_{x}$ like
 $\bar{z}_x\partial z_x/\partial \tau, 
 \bar{\psi}_x\partial \psi_x/\partial \tau$ disappear, and 
 the $T$-dependence may appear only 
 as an overall factor
$\beta$, which may be absorbed into the coefficients of the action 
and one may still deal with the 3D model
instead of the four-dimensional model.
Study of finite-$T$ properties of the systems gives us an
important insight into the low-$T$ phase structure, for
we can expect that ordered phase at finite-$T$ generally survives 
at $T=0$.

In this way, the partition function $Z$ of the 3D model at finite $T$'s 
is given by the path integral\cite{im},
\be
Z&=&\int [dz][d\psi][dU]\exp A, \label{Z} \\
\left[dz\right]&=&\prod_{x}dz_x,\ \left[d\psi\right]=\prod_x d\psi_x
d\bar{\psi}_x,\
\left[dU\right]=\prod_{x,\mu} dU_{x\mu},
\nonumber
\ee
with the following action $A$ on the 3D lattice\cite{star,c2}, 
\be
A&=&A_{\rm AF}+A_{\rm hop}+A_{\rm SC},\nn
A_{\rm AF}&=&\frac{c_1}{2}\sum_{x,\mu}
\Big(z^\star_{x+\mu}U_{x\mu}z_{x} + \mbox{c.c.}\Big), \nn
A_{\rm hop}&=&{c_3 \over 2}
\sum_{x,\mu}\left(\bar{z}_{x+\mu}z_x 
\bar{\psi}_x\psi_{x+\mu} +{\rm c.c.}\right)-m\sum_x \rho_x,
\nn
A_{\rm SC}&=&{J\beta \over 2}\sum_{x,\mu}\rho_{x+\mu} \rho_x
|z^\star_{x+\mu}z_{x}|^2,
\label{action}
\ee
where
\be
U_{x\mu} &\equiv& \exp(i\theta_{x\mu})\ \in U(1),\nn
\rho_x&\equiv&\bar{\psi}_x\psi_x,\nn
\bar{z}_{x+\mu}z_x&\equiv&\bar{z}_{x+\mu,1} z_{x1}+
\bar{z}_{x+\mu,2} z_{x2}\nn
z^\star_{x1}&\equiv&z_{x2},\ z^\star_{x2}\equiv -z_{x1},\nn
z^\star_{x+\mu}z_x&=&z_{x+\mu,2}z_{x1}-z_{x+\mu,2}z_{x1}.
\label{action2}
\ee
The first term $A_{\rm AF}$ in the action $A$ 
describes the AF coupling between NN spinons.
We have introduced the  U(1) gauge
field $U_{x\mu}$ on the link $(x,x+\mu)$
as an auxiliary field to make the action in a simpler form
and the U(1) gauge invariance (explained below) manifest.
The second term $A_{\rm hop}$ describes
simultaneous NN hopping of  a holon and a spinon keeping 
its spin orientation (i.e., in the FM channel).
The third term $A_{\rm SC}$ describes
attractive force between hole pairs, which we shall discuss 
in Sect.IID in detail.
There are remaining terms\cite{irrelevant}, 
which are irrelevant to discuss the global phase structure.
  
The integration measure of $z_{x\sigma}$ and $U_{x\mu}$ are
\be
\int dz_x &=&\prod_{\sigma}\int_{-\infty}^\infty\!\!\!\! 
d\,{\rm Re}z_{x\sigma}
\int_{-\infty}^\infty\!\!\!\! d\,{\rm Im}z_{x\sigma}\cdot
 \delta(\sum_\sigma\bar{z}_{x\sigma}z_{x\sigma}-1),\nn
\int dU_{x\mu}&=& \int_{-\pi}^{\pi}\frac{d\theta_{x\mu}}{2\pi}.
\ee
Grassmann variables $\psi_x$ anti-commute each other,
\be
[\psi_x, \psi_{x'} ]_+ =[\psi_x, \bar{\psi}_{x'} ]_+ =
[\bar{\psi}_x, \bar{\psi}_{x'} ]_+ = 0.
\ee
The formulae of  Grassmann integration\cite{berezin} are
\be
\int d\psi_xd\bar{\psi}_x [1,\psi_x,\bar{\psi}_x,\bar{\psi}_x\psi_x]=[0,0,0,1].
\ee

The term $m\sum_x\bar{\psi}_x\psi_x$ adjust the hole density to $\delta$
as
\be
\la \bar{\psi}_x\psi_x \ra =\delta.
\label{delta}
\ee
Therefore the parameter $m$ works as (the minus of) the chemical potential.

The action $A$ is invariant 
under a local ($x$-dependent) U(1) gauge transformation
with a gauge function $\lambda_x$\cite{utrsf},
\be
z_{x\sigma} &\rightarrow& e^{i\lambda_x}z_{x\sigma}, \nn 
U_{x\mu}&\rightarrow& e^{-i\lambda_{x+\mu}}U_{x\mu}e^{-i\lambda_x},\nn 
\psi_x &\rightarrow& e^{i\lambda_x} \psi_x.
\label{gaugetr}
\ee

\subsection{AF and Ferromagnetic spinon amplitudes}

The gauge field $U_{x\mu}$ is related to the spinon field $z_x$ as 
\be
\la U_{x\mu} \ra \sim \overline{{\left\la
\frac{{z}^\star_{x+\mu} z_{x}}{|{z}^\star_{x+\mu} z_{x}|}\right\ra}},
\label{UAF}
\ee
which is obtained by maximizing the action $A_{\rm AF}$.
Therefore $U_{x\mu}$ describes the (c.c.\,of) phase factor of 
AF NN spin-pair amplitude ${z}^\star_{x+\mu} z_{x}$
of Eq.(\ref{action2}).
In fact, one can integrate out $U_{x\mu}$ in  Eq.(\ref{Z}) and obtain
\be
&& \int [dU]\exp(A_{\rm AF}) = \exp(\tilde{A}_{\rm CP^1}),\nn
&& \tilde{A}_{\rm CP^1}=\sum_{x\mu}\log I_0(c_1|z^\star_{x+\mu}z_{x}|),
\label{I0}
\ee
where $I_0$ is the modified Bessel function.
The effective term $\tilde{A}_{\rm CP^1}$
should be compared with the original expression
$A_{\rm CP^1}$ of the CP$^1$ model,
\be
Z_{\rm CP^1} &=& \int [dz]\exp(A_{\rm CP^1}),\nn
A_{\rm CP^1} &=& \frac{\beta J}{2}\sum_{x,\mu}\big
|z^\star_{x+\mu}z_{x}\big|^2.
\ee
This CP$^1$ model describes the $t$-$J$ model without holes
($c_3=0$, $A_{\rm SC}=0$), i.e.,
AF Heisenberg spin model at finite $T$. 
Note that the  amplitude $z^\star_{x+\mu}z_x$ 
between NN spinon pair reads explicitly as
\be 
z^\star_{x+\mu}z_x =z_{x+\mu,2}z_{x1}-z_{x+\mu,1}z_{x2}.
\ee
This expresses the amplitude of spin-singlet AF combination
of NN spinons, which is called the RVB.
Both models with $A_{\rm CP^1}$ and $\tilde{A}_{\rm CP^1}$
 have similar behavior and it is verified 
that they  give rise to 
second-order transitions at certain $c_1$ and $J$\cite{CPN-2}.
The parameters $c_1$  in the action (\ref{action})
are related to the original ones as\cite{CPN-2}
\be
c_1&\sim& 
\left\{
\begin{array}{ll}
J\beta& {\rm for}\ c_1 \gg 1,\\
(2J\beta)^{1/2}& {\rm for}\ c_1 \ll 1,
\end{array}
\right.
\label{c1c3}
\ee
For the coupling $c_3$, the relation is straightforward,
\be
c_3&\sim& t\beta.
\ee

Let us see the meaning of CP$^1$ term $A_{\rm AF}$ (the AF spin coupling) 
and the hopping term $A_{\rm hop}$ (the $t$-term) further.
For this purpose, it is convenient to introduce
an O(3) spin vector field $\vec{\ell}_x$ made of spinon $z_x$,
\be
\vec{\ell}_x \equiv \bar{z}_x\vec{\sigma}z_x =
\sum_{\sigma,\sigma'}\bar{z}_{x\sigma}\vec{\sigma}_{\sigma\sigma'}
z_{x\sigma'},\quad 
\vec{\ell}_x\cdot\vec{\ell}_x=1.\quad
\label{ell}
\ee
The NN spin correlation
$\vec{\ell}_{x+\mu}\cdot\vec{\ell}_x$ is expressed by the 
CP$^1$ amplitudes (such as $\bar{z}_{x+\mu}z_{x}$) as
\be
\vec{\ell}_{x+\mu}\cdot\vec{\ell}_x &=& 2|\bar{z}_{x+\mu}z_{x}|^2 -1 \nn
&=& -2|z^\star_{x+\mu}z_{x}|^2 +1,
\ee
where we have used the identity, 
\be
|\bar{z}_{x+\mu}z_x|^2+|z^\star_{x+\mu}z_x|^2=1.
\label{sumrule}
\ee
So if the spinon hopping amplitude $\bar{z}_{x+\mu}z_x$
which appears in $A_{\rm hop}$ has an absolute value near its maximum,
$|\bar{z}_{x+\mu}z_x|\sim 1$, then the NN spins
are mostly FM $\vec{\ell}_{x+\mu}\cdot\vec{\ell}_x\sim 1$.
On the other hand, if  the spinon RVB amplitude
$z^\star_{x+\mu}z_x$
takes  values with $|z^\star_{x+\mu}z_x|\sim 1$, then
the NN spins are mostly  AF, $\vec{\ell}_{x+\mu}\cdot\vec{\ell}_x\sim -1$.
These two amplitudes satisfy the sum rule (\ref{sumrule}).
The AF phase and FM phase are characterized by the LRO
in the spin correlation function 
$\la \vec{\ell}_x \cdot\vec{\ell}_y \ra$,
and they can coexist with each other as we see in the following
sections.

\subsection{Holon hopping expansion and auxiliary field $V_{x\mu}$}

In Eq.(\ref{Z}), one can integrate out 
the fermionic holon field $\psi_x$ by assuming small holon density 
$\delta$ of Eq.(\ref{delta}).
In this region, the hopping expansion of $\psi_x$ is applicable as 
it is an expansion in powers of $\delta$.
Some details of the integration of $\psi_x$ are given in Appendix A.
After integration over $\psi_x$ we obtain
\be
&&\int[d\psi]\exp(A_{\rm hop})= \exp(\tilde{A}_{\rm hop}),\nn
&&\tilde{A}_{\rm hop} = \delta \Big({c_3 \over 2}\Big)^2\sum_{x,\mu}
 \big|\bar{z}_{x+\mu}z_x\big|^2 \nn
&&\hspace{0.8cm}+\delta \Big({c_3 \over 2}\Big)^4\sum_{x, \mu < \nu}
\prod_{{\rm plaq.}}(\bar{z}_{x+\mu}z_x)+\cdots.
\label{hopping}
\ee
The second term of $\tilde{A}_{\rm hop}$ denote the product of 
$\bar{z}_{x+\mu}z_x$ on the link $(x,x+\mu)$
[$\bar{z}_xz_{x+\mu}$ on the link $(x+\mu,x)$]
around the plaquette $(x, x+\mu, x+\mu+\nu, x+\nu)$ 
and the ellipsis denotes non-local higher-order terms.
Both the first and the second terms favor FM couplings of NN spin
pairs.

Then we introduce a vector field $W_{x\mu}$ as an auxiliary
field corresponding to
$\bar{z}_x{z}_{x+\mu}$,
\be
\la W_{x\mu} \ra \sim \la \bar{z}_x z_{x+\mu}\ra, 
\label{WFM}
\ee
by using Gaussian integration (Hubbard-Stratonovich
transformation) as follows,
\be
\hspace{-0.5cm}
\exp(\tilde{A}_{\rm hop})&=&\int [dW]\exp(A_W),\nn
A_W&=& \delta \Big({c_3 \over 2}\Big)^2\Big(-\sum_{\rm x,\mu}|W_{x\mu}|^2 \nn
&&+\sum_{x,\mu}\left(\bar{z}_{x+\mu}z_xW_{x\mu}
+\mbox{c.c.}\right)\Big) \nn
&&+\delta \Big({c_3 \over 2}\Big)^4\sum_{x, \mu < \nu}
\prod_{{\rm plaq.}}W_{x\mu}+\cdots.
\label{hopping2}
\ee
Estimation of the magnitude of $W_{x\mu}$ is straightforward for 
$T/J\alt 1$ as 
\be
\int [dz]e^{-\frac{J}{2}\beta |\bar{z}_{x+\mu}z_x|^2}
|\bar{z}_{x+\mu}z_x|^2\sim \frac{1}{J\beta}.
\ee
Then we set 
\be
W_{x\mu}&=& W V_{x\mu},\quad W \simeq \frac{1}{\sqrt{J\beta}},\nn
V_{x\mu}&=& \exp(i\phi_{x\mu})\ \in {\rm U(1)},
\label{ampW}
\ee
by ignoring the fluctuation of radial component of $W_{x\mu}$
and focusing on its phase,
$dW_{x\mu} \to dV_{x\mu} \equiv d\phi_{x\mu}/(2\pi)$.
So the correspondence (\ref{WFM}) becomes
\be
\la V_{x\mu} \ra \sim \Big\la \frac{\bar{z}_x z_{x+\mu}}{|\bar{z}_x z_{x+\mu}|}
\Big\ra. 
\ee
This simplification is based on the observation that 
the most relevant degrees of freedom in gauge theories
are the phases of gauge fields on the links rather than
the amplitude $W$ because the latter has only massive excitations. 
The phase factor $V_{x\mu}$ is a new U(1) gauge field that transforms 
under the gauge transformation (\ref{gaugetr}) as
\be
V_{x\mu} \rightarrow e^{i\lambda_{x+\mu}}
                     V_{x\mu}e^{-i\lambda_x}.
\ee
Physical meaning of $V_{x\mu}$ is obvious from the discussion 
given in Sect.IIB.
It measures the phase part of SR FM spinon channel,
 the deviation from the AF order.
Its coherent ``condensation" 
induces coherent hopping of holons $\psi_x$
in the FM spinon channel as the term $A_{\rm hop}$ shows,
and therefore a MI transition into a metallic phase.
More detailed discussion will be given in the following sections.

The $A_{\rm hop}$ term  is then rewritten effectively as follows,
\begin{eqnarray}
\hspace{-1cm}
\exp(\tilde{A}_{\rm hop}) &=& \int [dV] \exp(A_V),\nn
A_V &=& \frac{c_4}{2}\sum_{x,\mu}\left(
V_{x\mu}\bar{z}_{x+\mu}z_{x}+\mbox{c.c.} \right)\nn
&+&\frac{c_5}{2}\sum_{x,\mu<\nu}
\left( \bar{V}_{x\nu}\bar{V}_{x+\nu,\mu}V_{x+\mu,\nu}V_{x\mu}
+\mbox{c.c.} \right),\nn
\int[dV]&=&\prod_{x,\mu}\int_{-\pi}^{\pi} \frac{d\phi_{x\mu}}{2\pi}.
\label{avav}
\end{eqnarray}
We have neglected the higher-order terms in Eq.(\ref{hopping}) as
they have smaller coefficients for $T/J<1$ with numerical damping factors.
However, effects of these non-local terms can be expected qualitatively.
As they have all positive coefficients, all of them favor the order of the field $V_{x\mu}$ and so the metallic phase. 
From this point of view, the critical hole concentration $\delta_c$ of
the MI transition 
obtained by the numerical study in Sect.IV might give an 
overestimation for the true value. 

The parameters $c_4$ and $c_5$ in $A_V$ are related to the original ones as
\be
&& c_4\sim \frac{\delta c^2_3}{J\beta}\sim 
\frac{\delta t^2\beta}{J}, \nn
&& c_5\sim \frac{\delta c^4_3}{(J\beta)^2}\sim 
\frac{\delta t^4 \beta^2}{J^2}.
\label{c4c5}
\ee
In the following investigation of the phase diagram of the system, 
however, we treat 
$c_4$ and $c_5$ in more flexible manner 
as free parameters that are proportional to $\delta$
and are increasing functions of $\beta=1/T$.
As most of phase transitions in the present model appear in the region
$c_1\gg 1$, we identify $T$ and $\delta$ from Eqs.(\ref{c1c3}, \ref{c4c5}) as
\be
T \simeq \frac{J}{c_1},\quad 
\delta \simeq \frac{J^2}{t^2}\frac{c_4}{c_1}.
\label{c1c4tdelta}
\ee

At this stage, the original partition function $Z$ 
without $A_M$ is expressed as
\be
Z \to Z_{UV} &\equiv& \int [dz][dU][dV] \exp(A_{UV}),\nn
A_{UV} &=& A_{\rm AF}(z_x,U_{x\mu}) + A_V(z_x,V_{x\mu}).
\label{AUV}
\ee
This ``UV" model describes the competition between the
AF-RVB spin-pair amplitude $U_{x\mu}$ and the
FM spin-hopping amplitude $V_{x\mu}$, the latter is generated 
by integration over holon hopping.

\subsection{Hole-pair field $M_{x\mu}$ and the full model $A_{\rm full}$}

As shown in the previous section, in the effective action
(\ref{action}),
there exists the term $A_{\rm SC}$ that describes
an attractive force between NN holes doped in AF magnets,
or more precisely, in a SR AF background.
This attractive force comes from the $J$-terms in the Hamiltonian
(\ref{htj2}). 
Actually, the two holes with a mutual distance more than 
one lattice spacing break twelve AF bonds of spins,
while a pair of holes at NN sites break just eleven AF bonds.
Thus the NN hole pair is favored energetically.   
To see it explicitly, we rewrite $A_{\rm SC}$ in Eq.(\ref{action})
as follows,
\begin{equation}
A_{\rm SC}={J\beta \over 2}\sum_{x,\mu} \Big|\bar{\psi}_{x+\mu}
\big(z^\star_{x+\mu}z_{x}\big)
\bar{\psi}_{x}\Big|^2.
\label{Asc}
\end{equation}
Note that the holon-pair variable $\bar{\psi}_{x+\mu}\bar{\psi}_{x}$
 is accompanied with the RVB spinon-pair amplitude
 $z^\star_{x+\mu}z_{x}$.
 This combination is nothing but $C_{x+\mu,2}C_{x1}-
 C_{x+\mu,1}C_{x2}$ in terms of electron operators.
We expect that this attractive force induces hole-pair condensation
under certain conditions, and as a result, a  SC state is generated.
The main problem to be addressed here is whether 
the above attractive force is 
strong enough to generate a SC state in the region {\em without} AF
LRO.

In order to investigate a possible SC phase transition, 
we introduce a hole-pair field
$M_{x\mu}$ as a complex auxiliary field
 describing the configuration of holon-pair accompanied with
 the RVB spinon pair  at the sites $x$ and $x+\mu$.
So  $M_{x\mu}$ should satisfy
\be
\la M_{x\mu} \ra \sim 
\la \bar{\psi}_{x+\mu}\big(z^\star_{x+\mu}z_{x}\big)
\bar{\psi}_{x}\ra.
\label{HP}
\ee
This hole-pair field $M_{x\mu}$ is nothing but annihilation operator of 
spin-singlet electron pair sitting NN sites as mentioned.
Explicitly, we use the Hubbard-Stratonovich
transformation for $A_{\rm SC}$ as Eq.(\ref{hopping2}),
\be
&&\exp\left({J \over 4} \Big|\bar{\psi}_{x+\mu}
\big(z^\star_{x+\mu}
{z}_{x}\big)\bar{\psi}_{x}\Big|^2\right) \nn
&=&\int dM_{x\mu} 
\exp \Big[-{J\beta \over 4}\bar{M}_{x\mu} M_{x\mu}  \nn
&&+{J\beta \over 4}\Big(M_{x\mu}\psi_{x}
\big(\bar{z}^\star_{x+\mu} \bar{z}_{x}\big)
\psi_{x+\mu}+\mbox{c.c.}\Big)\Big].
\label{ascm}
\ee
This assures us of Eq.(\ref{HP}).

To study the effect of $A_{\rm SC}$, we start with $Z$ of Eq.(\ref{action}) 
and rewrite $A_{\rm SC}$ in the action by using Eq.(\ref{ascm}).
Then we integrate out the holon variables $\psi_x$ as in the previous
case (without $A_{\rm SC}$ there) to obtain the effective action, 
$A_{\rm full}$, where the suffix ``full" implies $A_{\rm SC}$
is taken into account. 
The partition function of the full model 
is now given as
\be
Z_{\rm full}&\equiv& \int [dz][dU][dV][dM]\exp(A_{\rm full}),\nn
A_{\rm full} &=& A_{UV}+A_M = A_{\rm AF} + A_V + A_M.
\label{Afull}
\ee 
In addition to the action of the $UV$ model of Eq.(\ref{AUV}),
$A_{\rm full}$ includes an extra  term $A_{\rm M}(z_x, M_{x\mu})$ 
that depends  on $z_x$ and $M_{x\mu}$. 

We have calculated $A_M$ in the order up to $O((c_3)^4)$\cite{im}.
Eq.(\ref{ascm}) shows that
as $\psi_x$ and $\psi_{x+\mu}$ hop, they leave 
the factor $M_{x\mu}\bar{z}^\star_{x+\mu} \bar{z}_{x}$, i.e.,
$M_{x\mu}$ is always accompanied with
the AF component of spinon, (c.c. of) ${z}^\star_{x+\mu}z_x$.
The hopping term $A_{{\rm hop}}$ itself supplies the FM component
$\bar{z}_{x}z_{x+\nu} (\sim |\bar{z}_{x}z_{x+\nu}|V_{x\nu})$  
along the link $(x,x+\nu)$ $\psi_x$ hops.
In expressing $A_M$ we prefer to use $U_{x\mu}$ 
instead of $\bar{z}^\star_{x+\mu}\bar{z}_x$ using Eq.(\ref{UAF}),
because it makes the gauge invariance of the system manifest. 
Then $M_{x\mu}$ appears in $A_M$ in the combination
$M_{x\mu}(\bar{z}^\star_{x+\mu}\bar{z}_x)
 \sim M_{x\mu}|z^\star_{x+\mu}z_x|U_{x\mu}$.
So we define a new variable,
\be
M_{x\mu}^\star \equiv M_{x\mu}U_{x\mu}\sim \bar{\psi}_{x+\mu}\bar{\psi}_x.
\ee
and write $A_M$ in terms of $M^\star_{x\mu}$ and $V_{x\mu}$,
the latter is supplied by $A_{{\rm hop}}$.
We note that $M_{x\mu}^\star$ is not gauge-invariant and represents
the ``holon pair" at $(x,x+\mu)$ in contrast with gauge-invariant $M_{x\mu}$
for ``hole pair".

In the practical calculations in Sect.IV,
we focus on the phase degrees of freedom of $M^\star_{x\mu}$, 
ignoring fluctuations of the 
radial part of $M^\star_{x\mu}$ as in the case of $W_{x\mu}\to V_{x\mu}$
(the London limit).
So we set
\be
M^\star_{x\mu} &=&M \exp(i\varphi_{x\mu}),\nn
M &\simeq& \sqrt{\mbox{holon-pair density}} \sim \delta,
\label{ampM}
\ee
and $dM_{x\mu} = d\varphi_{x\mu}/(2\pi)$.
We include $M$ into 
the coefficients of $A_{\rm M}$ and treat $M_{x\mu}=\exp(i\varphi_{x\mu})$
as a U(1) variable.
Furthermore, 
we regard $|\bar{z}_x z_{x+\mu}|$ and $|z^\star_{x+\mu}z_x|$ 
involved in $A_M$ as constants. They are also absorbed in the coefficients.
The reason of this treatment is given below on the determination of 
the coefficients.

\begin{figure}[t]
\begin{center}
\includegraphics[width=0.6\hsize]{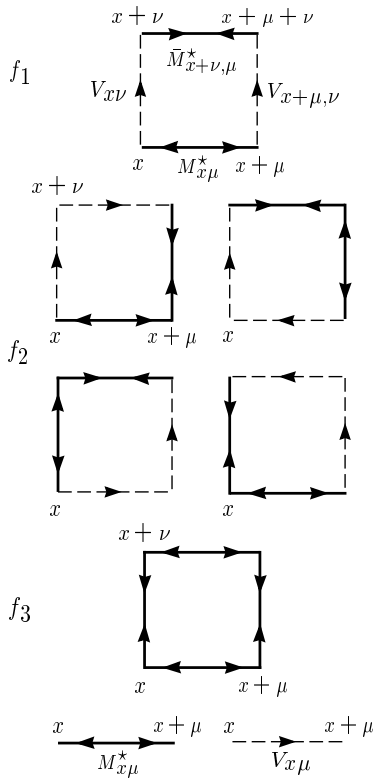}
\caption{Each term of $A_{\rm M}$ of Eq.(\ref{AM}).
The lines with the reversed arrows indicate
complex-conjugate variables, $\bar{M}^\star_{x\mu},\ \bar{V}_{x\mu}$. 
The gauge invariance under Eq.(\ref{gaugetr}) forces 
the arrows near each corner to make a divergenceless flow.
}
\label{fig1}
\end{center}
\end{figure}

In terms of this $M_{x\mu}^\star$, $A_M$ is expressed as 
\be
A_{M} &=& \frac{f_1}{2}
\sum_{x,\mu\neq\nu}M^\star_{x\mu}\bar{M}^\star_{x+\nu,\mu}
V_{x+\mu,\nu}V_{x\nu}\nonumber \\
&& +\frac{f_2}{2}\sum_{x,\mu<\nu} \alpha_{\mu\nu}
\Big[V_{x\nu}V_{x+\nu,\mu}\bar{M}^\star_{x+\mu,\nu}M^\star_{x\mu}
 \nonumber\\
&&\hspace{2cm}+V_{x\nu}\bar{M}^\star_{x+\nu,\mu}
M_{x+\mu,\nu}^\star\bar{V}_{x\mu}  \nonumber\\
&&\hspace{2cm}+M^\star_{x\nu}\bar{M}^\star_{x+\nu,\mu}V_{x+\mu,\nu}V_{x\mu}
 \nonumber \\
&&\hspace{2cm}+\bar{M}^\star_{x\mu}
\bar{V}_{x+\nu,\mu}V_{x+\mu,\nu}M_{x\mu}^\star \Big] \nn
&& +\frac{f_3}{2}\sum_{x,\mu<\nu}\bar{M}^\star_{x\nu}
M^\star_{x+\nu,\mu}\bar{M}^\star_{x+\mu,\nu}{M}^\star_{x\mu} \nn
&&+\mbox{c.c.}.
\label{AM}
\ee
Each term in $A_{M}$ is schematically shown in Fig.\ref{fig1}.
The terms with the coefficients $f_1$ and $f_2$ in Eq.(\ref{AM})
describe the local hopping of the holon-pair field $M^\star_{x\mu}$,
whereas the $f_3$-term controls fluxes of $M^\star_{x\mu}$ penetrating
each plaquette. These fluxes 
correspond to vortex excitations in the SC state.
In other words, the $f_1$ and $f_2$-terms induce a primordial SC state and
a genuine SC state is generated by the $f_3$-term.
Numerical investigations in the following sections verify this
qualitative expectation.

As stated in Sect.I, we think that a SC state is to be realized
in a metallic phase, i.e., beyond the region of applicability of
the leading order of the hopping expansion. 
So keeping the results of the hopping expansion
for the three coefficients $f_i$ of $A_M$  is not suitable for
discussing a SC state.
For example, the coefficient $f_3$ is negative in the leading order
of the hopping expansion, which favors the $s+id$-wave SC. 
We examined higher-order terms of the hopping expansion and found that
some of them generate a positive value of $f_3$ to support
the $d$-wave SC as observed experimentally. 
So in the following numerical studies, we shall assume that 
$f_i$'s are positive and proportional to $\delta^2$,    
\be
f_1,f_2,f_3\propto \delta^2.
\ee 
and treat their coefficients as {\it positive} free
(phenomenological) parameters\cite{FNf3}.
In short, all the effect of $|M_{x\mu}|$, $|\bar{z}_x z_{x+\mu}|$ and 
$|z^\star_{x+\mu}z_x|$ in and near the SC state are included in the 
effective coefficients $f_i$ of $A_M$.

Also we have incorporated in Eq.(\ref{AM})  
the  layered-structure of the 3D lattice
of cuprates\cite{3dasymmetry} by introducing in Eq.(\ref{AM})
the anisotropy parameter $\alpha_{\mu\nu}$, which is defined as 
\begin{equation}
\displaystyle\alpha_{\mu\nu}=\displaystyle\alpha_{\nu\mu}=\left\{
\begin{array}{ll}
1 &\quad (\mu,\nu\neq 3) \\
0 &\quad (\mu\;\mbox{or}\;\nu=3)
\end{array}
\right.\ .
\label{alpha}
\end{equation}
The layered structure of the system is systematically incorporated
in the original Hamiltonian (\ref{htj}) by making the parameters $t$
and $J$ anisotropic.
This induces anisotropies in the effective model that we have derived.
Most of the terms are insensitive to the anisotropy except the $f_2$-term,
which is the reason that we treat $A_{{\rm hop}}$ and $A_V$
in a symmetric manner.
For the $f_2$-term, the layered structure plays an important role
to avoid frustrations and make the symmetry of SC to be 
$d_{{\rm x}^2-{\rm y}^2}$.

\section{Phase structure of the $UV$ model: AF and MI transitions}
\setcounter{equation}{0}

In this section, we study the $UV$ model with
the action $A_{UV} =A_{\rm AF}+A_V$ in Eq.(\ref{AUV}) 
by means of the MC simulations.
The full model $A_{{\rm full}}=A_{\rm AF}+A_V+A_M$ 
shall be studied in Sect.IV.
For MC simulations, we consider a 3D cubic lattice of the size 
$V\equiv L^3$ ($L$ up to 30) with the periodic boundary condition.
We used the standard Metropolis algorithm for local update.
Average number of sweeps was $2\times 10^{5}$, and acceptance ratio was 
about $\sim 40\%$. 

To study the phase structure of the model, we measured the internal energy 
$E$ and the specific heat $C$, which are defined as 
\be
E=-\frac{1}{L^3}\langle A \rangle, \quad
C=\frac{1}{L^3}\langle (A-\langle A \rangle)^2\rangle,
\label{EC}
\ee
as functions of the parameters $c_1$, $c_4$ and $c_5$.
We note that the $c_4$ term and $c_5$ term are 
related with each other because both are generated by $c_3$ term.
Below we respect this correlation by setting
the parameter $c_5$ as $c_5\propto c_4$.

By obtaining the locations of the phase transition lines
by the peaks of $C$, etc., we get a phase diagram in the 
$c_4-c_1$ plane.
Then we investigate spin correlation functions
and instanton densities in order to
identify the physical meaning and properties of each phase.
To support this procedure, we also investigated 
fluctuations of each term of the action by measuring
the individual ``specific heat" $C_{Ai}$ defined by
\be
C_{A_i}&\equiv& \frac{1}{L^3}
\langle (A_{i}-\langle A_{i}\rangle)^2\rangle,\quad i=1,4,5,\nn
A_1&\equiv& A_{\rm AF},\ A_{4,5} \equiv c_{4,5}{\rm -term\ in}\ A_V. 
\label{partialc}
\ee

At $c_4=0$ (i.e., at $c_3=0$), the system  is reduced to the
AF Heisenberg model with the action $A_{\rm AF}$ alone,
which has a phase transition
from the paramagnetic (PM) spin-disordered phase to the AF  spin-ordered
phase at $c_1\sim 2.8$.
We study how the location of this AF phase transition changes 
and whether new phases appear as the $c_4$-term is turned on.
It is naturally expected that the AF phase transition shifts to
low-$T$ region (large $c_1$ region) as the parameter $c_4$ is 
increased because the $c_4$ term favors FM NN spin coupling.

\begin{figure}[t]
\begin{center}
\includegraphics[width=0.6\hsize]{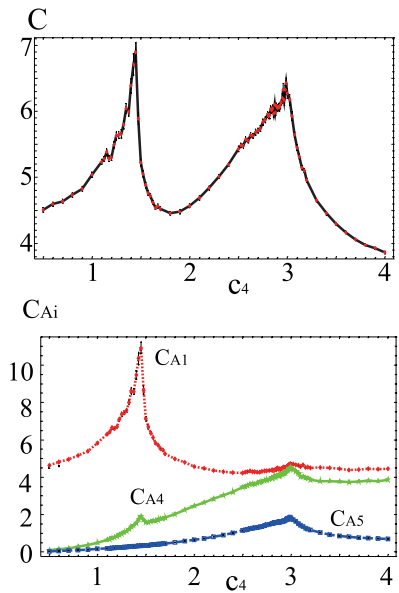}
\caption{Total specific heat $C$ and specific heat of each term 
$C_{A_i}$
of Eq.(\ref{partialc}) as functions of $c_4$ for $c_1=3.5$
and $c_5 = c_4/3.0$.
System size is $L=30$.
$C$ has two peaks at $c_4 \simeq 1.5, 3.0$.
$C_{A_1}$ has  a sharp peak at $c_4 \simeq 1.5$
 suggesting the AF transition, and
$C_{A_4}, C_{A_5}$ have peaks at $c_4 \simeq 3.0$ suggesting
the MI transition.}
\label{fig2}
\end{center}
\end{figure} 

Let us  first examine $C$ and $C_{A_i}$ as  functions of $c_4$ for $c_1=3.5$.
As we shall see later on, this value of $c_1$ belongs to
relatively high-$T$ region.
In Fig.\ref{fig2} we show the result for the case $c_5=c_4/3.0$.
We found no anomalous behavior of $E$ such as hysteresis, 
whereas $C$ shown in Fig.\ref{fig2} exhibits two sharp peaks at 
$c_4 \simeq 1.5$ and $3.0$.
We verified that each peak has a systematic system-size ($L$) dependence,
so we concluded that both peaks show existence of second-order 
phase transitions.
$C_{A_1}$ of Fig.\ref{fig2} exhibits a very sharp peak at $c_4\simeq 1.5$.
On the other hand, both $C_{A_4}$ and $C_{A_5}$ exhibit a peak at 
$c_4\simeq 3.0$.
Then we conclude that the AF phase transition takes place at
$c_4\simeq 1.5$ and the MI transition at
$c_4\simeq 3.0$.

\begin{figure}[tbp]
\begin{center}
\includegraphics[width=0.7\hsize]{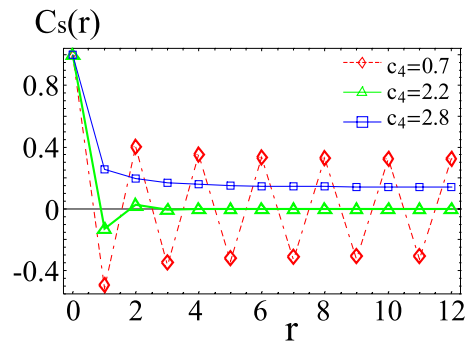}
\caption{Spin correlation function $G_s(r)$ of Eq.(\ref{SC}) for 
$c_1=3.5,\ c_5 = c_4/3.0$ and $L=24$. 
At $c_4=0.7$, an AF LRO exists.
At $c_4=2.2$, the AF LRO disappears.
At $c_4=3.8$, a FM correlation appears as a result of existence of 
``free electrons".
}
\label{fig3}
\end{center}
\end{figure}

The above conclusion may be confirmed by calculating
the spin correlation function. 
In Fig.\ref{fig3}, we show the correlation function
$G_{\rm s}(r)$ of the O(3) spin $\vec{\ell}_x$ of Eq.(\ref{ell}),
\be
G_{\rm s}(r)=\frac{1}{3L^3}\sum_{x,\mu}\langle \vec{\ell}_x\cdot
\vec{\ell}_{x+r\mu}\rangle.
\label{SC}
\ee
As we expected, at $c_4=0.7$, 
$G_{\rm s}(r)$ exhibits an oscillatory behavior and
has a staggered magnetization,
\be
&&\lim_{r\to \infty}(-)^r G_{\rm s}(r)
\simeq (-)^{r_{\rm max}} G_{\rm s}(r_{\rm max}) \neq 0,\nn
&& r_{\rm max} \equiv\frac{L}{2},\quad ({\rm AF\ phase}).
\ee
So there exists an   AF LRO at $c_4=0.7$.
This confirms that the phase transition at $c_4\simeq 1.5$ is the 
AF transition. 
At $c_4=2.2$ this AF order disappears and the system
is in a magnetically disordered phase that we call
paramagnetic (PM) phase.
At $c_4=3.8$, $G_{\rm s}(r)$
exhibits a LRO,
\be
G_{\rm s}(r_{\rm max}) \neq 0,\quad ({\rm FM\ phase}), 
\ee
which implies that the system is in the FM phase.
So we obtain a picture of the phase structure for $c_1=3.5$ that,
as $c_4$ increases, the phase changes as
AF $\to$ PM $\to$ FM.

We also calculated instanton densities of the gauge fields $U_{x\mu}$
and $V_{x\mu}$.
For example, $U$-instanton density $\rho_U$ is defined for 
$U_{x\mu}=e^{i\theta_{x\mu}}, \ \theta_{x\mu}\in [-\pi,\pi]$
in the following way\cite{Inst-1,CPN-2}.
We first consider the magnetic flux $\Theta_{x\mu\nu}$ 
penetrating the plaquette $(x,x+\mu,x+\mu+\nu,x+\nu)$, which is defined as
\begin{eqnarray}
\Theta_{x\mu\nu}&\equiv&\theta_{x\mu}+\theta_{x+\mu,\nu}-\theta_{x+\nu,\mu}
-\theta_{x\nu}, \nonumber \\
 && (-4\pi\le \Theta_{x\mu\nu}\le 4\pi).
\label{Flux}
\end{eqnarray}
Then we decompose $\Theta_{x\mu\nu}$ into its integer part $n_{x\mu\nu}$,
which represents the Dirac string (vortex line), and the remaining 
fractional part $\tilde{\Theta}_{x\mu\nu}$,
\begin{equation}
\Theta_{x\mu\nu}=2\pi n_{x\mu\nu}+\tilde{\Theta}_{x\mu\nu}, \;\;
(-\pi\le \tilde{\Theta}_{x\mu\nu}\le \pi).
\label{Flux2}
\end{equation}
The $U$-instanton density $\rho_U(x)$ at the cube around the site
$x+{\hat{1} \over 2}+{\hat{2} \over 2}+{\hat{3} \over 2}$ of the
dual lattice is then defined as 
\begin{equation}
\begin{split}
\rho_U(x) &=-{1 \over 2}\sum_{\mu\nu\lambda}\epsilon_{\mu\nu\lambda}
(n_{x+\mu,\nu\lambda}-n_{x,\nu\lambda})  \\
&={1 \over 4\pi}\sum_{\mu,\nu,\lambda}\epsilon_{\mu\nu\lambda}
(\tilde{\Theta}_{x+\mu,\nu\lambda}-\tilde{\Theta}_{x,\nu\lambda}),
\end{split}
\label{rho}
\end{equation}
where $\epsilon_{\mu\nu\lambda}$ is the totally antisymmetric tensor.
From the above definition, we define the average instanton density 
$\rho_U$ as
\be
\rho_U\equiv \frac{1}{L^3}\sum_x
\la |\rho_U(x)|\ra.
\ee 
The $V$-instanton density $\rho_V$ is defined similarly for $V_{x\mu}$.

\begin{figure}[t]
\begin{center}
\includegraphics[width=0.70\hsize]{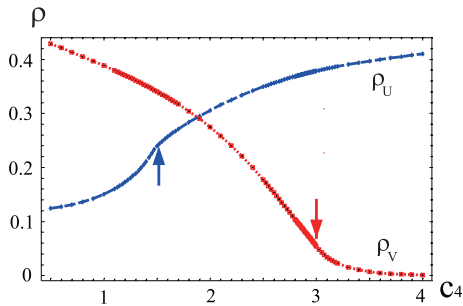}
\caption{Instanton densities $\rho_U$  and 
$\rho_V$ as functions of $c_4$ for $c_1=3.5,\ c_5 = c_4/3.0$ and $L=30$.
Arrows indicate the phase transition points
determined by the specific heat (See Fig.\ref{fig2}).
At the first phase transition point $c_4\simeq 1.5$, 
$\rho_U$ starts to increase.
On the other hand, at the second transition point $c_4\simeq 3.0$,
$\rho_V$ tends to vanish.
}
\label{fig4}
\end{center}
\end{figure} 

The instanton density $\rho_U$ measures strength of fluctuations
of the gauge field $U_{x\mu}$.
In the deconfinement phase of $U_{x\mu}$, 
fluctuations of $\Theta_{x\mu\nu}$ around its average 
$\Theta_{x\mu\nu} = 0$ are small and $\rho_U \simeq 0$.
In the confinement phase of $U_{x\mu}$, on the other hand,
$\Theta_{x\mu\nu}$  fluctuates violently, and
$\rho_U$ has a finite value.
Here we note that the confinement by $U_{x\mu}$ field
gives rise to  quasi-excitations that 
are gauge-invariant ``composite particles" in the AF channel.
Such combinations include
$z^\star_{x+\mu}U_{x\mu}z_x,\, 
\psi_{x+\mu}U_{x\mu}z_{x\sigma},$ etc. 
Similar interpretation holds for $\rho_V$ 
concerning to the gauge dynamics of
$V_{x\mu}$. 
The confinement here works in the FM channel,
and the possible gauge-invariant 
quasi-excitations are $\bar{\psi}_x z_{x\sigma}=C_{x\sigma},\, 
\bar{\psi}_x\psi_x,\, \bar{z}_{x\sigma}z_{x\sigma'}$
and their stretched versions such as  
$\bar{\psi}_{x+\mu}V_{x\mu}z_{x\sigma}$, etc.

In Fig.\ref{fig4} we show $\rho_U$ and $\ \rho_V$
for $c_1=3.5$. 
As we increases $c_4$, $\rho_U$ starts to increase
at the first phase transition at $c_4\simeq 1.5$.
This result means the fluctuation of $U_{x\mu}$ of AF NN spinon pairs 
become large, and the $U$-confinement spin-disordered phase appears.
This result is consistent with the interpretation based on $G_s(r)$ above. 
On the other hand, at the second phase transition at $c_4\simeq 3.0$,
$\rho_V$ tends to vanish.
So, for $c_4 < 3.0$, the system stays in the  $V$-confinement
phase  and holons and anti-spinons are 
bound within electrons as $\psi_x\bar{z}_{x\sigma}$.
For $c_4 > 3.0$, the system is in the $V$-deconfinement phase,
and holons and spinons start to hop coherently and independently as 
low-energy excitations. 
This indicates that the phenomenon of charge-spin 
separation\cite{css} takes place and also the system is metallic.

\begin{figure}[t]
\begin{center}
\includegraphics[width=0.6\hsize]{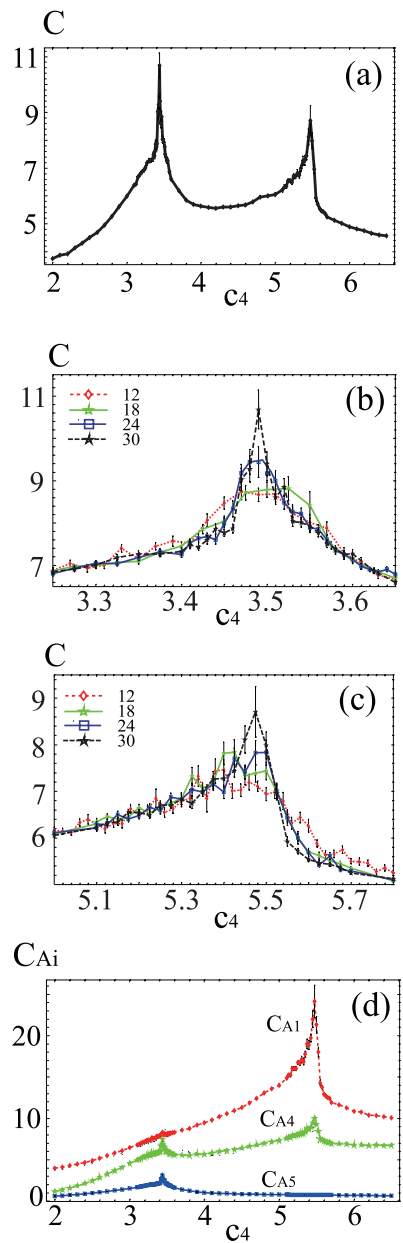}
\caption{Specific heat $C$ as a function of $c_4$ for $c_1=6.5$
and $c_5 = c_4/3.0$.
(a) $C$ for $L=30$. There are two peaks at $c_4 \simeq 3.4,\ 5.4$. 
(b,c) Each peak of $C$ develops as the system size is increased.
Results indicate that both phase transitions are of second order.
(d) Specific heat $C_{A_i}$ of each term. They indicate
the transition at $c_4 \simeq 3.4$ is the MI one and 
the transition at $c_4 \simeq 5.4$ is the AF one.}
\label{fig5}
\end{center}
\end{figure}

Let us turn to the low-$T$ region and see how the locations of these
 AF and MI phase transitions change.
In Fig.\ref{fig5}, we present the specific heat $C$ and
$C_{A_i}$ for $c_1=6.5$.
We again found two peaks at $c_4\simeq3.4$ and $5.4$.
Figs.\ref{fig5}b, c show that both peaks develops systematically
indicating that both phase transitions are of second order.
Individual specific heat in Fig.\ref{fig5}d shows
that the peak of $C$ at $c_4\simeq 3.4$ corresponds to
fluctuations of the $c_4$ and $c_5$-terms and so the MI transition, 
while the peak at $c_4\simeq 5.4$ is generated by the $c_1$-term and so
the AF transition.
Therefore the order of the AF and MI phase transitions along
the $c_4$ axis has been interchanged
compared to the previous high-$T$ case of $c_1=3.5$.

\begin{figure}[t]
\begin{center}
\includegraphics[width=0.7\hsize]{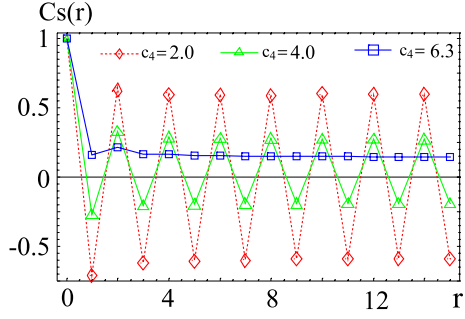}
\caption{Spin correlation function $G_s(r)$ for $c_1=6.5,\ c_5 = c_4/3.0$ 
and $L=30$.
At $c_4 = 2.0$, the oscillatory behavior around zero shows an AF order.
At $c_4 = 4.0$, there is an AF order in the FM background order.
At $c_4 = 6.3$, there is a FM order.
}
\label{fig6}
\end{center}
\end{figure}

To verify the above observation, we calculated the spin correlations,
$G_{\rm s}(r)$.
The result is shown in Fig.\ref{fig6}.
In the intermediate region $3.4 < c_4 <5.4$, $G_{\rm s}(r)$ exhibits
very interesting behavior, i.e., an AF correlation exists in a
FM background. This implies coexistence of the FM and AF orders. 
We note that a coexistence of FM and AM orders was also
observed previously in the CP$^1+$Higgs boson model\cite{btj}.
This model is a bosonic counterpart of the present model and
the U(1) Higgs variable $\exp(i\alpha_x)$ there plays the role
of the fermionic holon variable $\psi_x$.

\begin{figure}[b]
\begin{center}
\includegraphics[width=0.70\hsize]{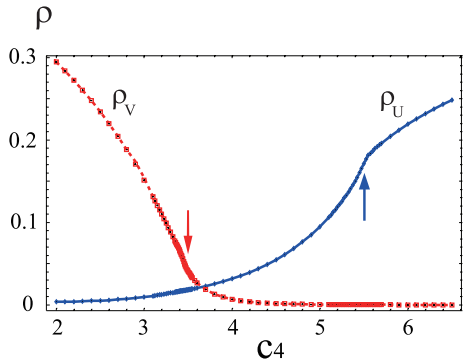}
\caption{Instanton densities $\rho_U$  and 
$\rho_V$ as functions of $c_4$ for $c_1=6.5,\ c_5 = c_4/3.0$ and $L=30$.
Arrows indicate the phase transition points
determined by the specific heat (See Fig.\ref{fig5}).
At the first transition point $c_4\simeq 3.4$, 
$\rho_V$ tends to vanish.
On the other hand, at the second transition $c_4\simeq 5.4$,
$\rho_U$ gets large values showing that the system enters into 
the $U$-confinement phase.
}
\label{fig7}
\end{center}
\end{figure}

In Fig.\ref{fig7} we present instanton densities.
Compared with the high-$T$ result of Fig.\ref{fig4},
the result again indicates that two phase transitions have interchanged
their order along the $c_4$ axis. As a result, 
there appears a range of $c_4$ in which both $\rho_U$ and $\rho_V$ are small,
which implies that spinons and holons hop here both in 
$U$ and $V$ channels.
In other words, charges are transported by holons whereas spin
degrees of freedom are transported by spinons both in 
the AF and FM channels.

\begin{figure}[t]
\begin{center}
\includegraphics[width=0.8\hsize]{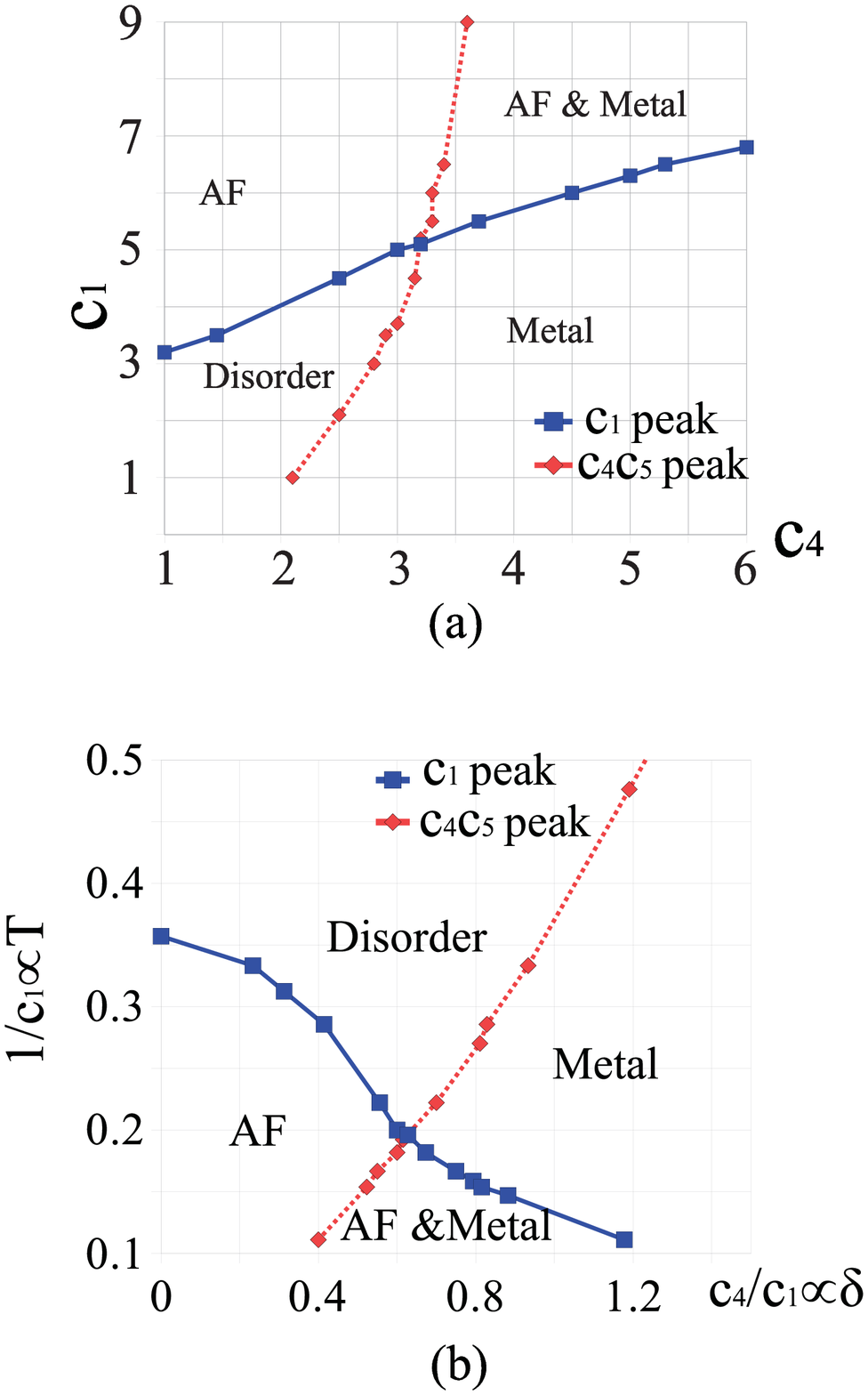}
\caption{Phase diagram of the UV model for $c_5 = c_4/3.0$.
(a) in the $c_4$-$c_1$ plane, and (b) in the $\delta$-$T$ plane.
Each phase is separated by two transition lines;
the AF transition line and MI transition line. 
All the transitions are of second order.}
\label{fig8}
\end{center}
\end{figure}

We repeated similar calculations for various values of $c_1$ and
$c_4$ and obtained the phase diagram of the model for $c_5 = c_4/3.0$.
In Figs.\ref{fig8}, we present  the phase diagram;
Fig.\ref{fig8}a in the $c_4-c_1$ plane and 
Fig.\ref{fig8}b in the $\delta-T$ plane, the latter is obtained
from the former by using Eq.(\ref{c1c4tdelta}).

So far, we have studied the case of $c_5/c_4 = 1/3.0$.
We also studied other values of the ratio 
$c_5/c_4$ and found similar phase diagram to that in
Figs.\ref{fig8}. 
As the value of $c_5/c_4$ is increased, the MI transition line
shifts to the region of smaller $\delta$.
This is expected from Eq.(\ref{avav}) 
because larger $c_5/c_4$ implies smaller critical value of $c_4$,
and therefore smaller critical $\delta$
from Eq.(\ref{c1c4tdelta}).

To estimate roughly the critical $\delta$ of the MI transition, 
which we call $\delta_{{\rm MI}}$, for real materials, 
one may put  $J\simeq 0.1{\rm eV},\ 
t \simeq 0.3{\rm eV}$ so $J/t\simeq 1/3$. 
Then we have $\delta \sim (J/t)^2\cdot (c_4/c_1) \simeq 
0.1(c_4/c_1)$ from  Eq.(\ref{c1c4tdelta}).
For $c_1\sim 10.0$ ($T\sim 100$K),
the critical line of Fig.\ref{fig8}a shows
$c_4/c_1\sim 0.4$, and this formula gives rise to
 $\delta_{{\rm MI}}\simeq 0.04$.
As discussed below Eq.(\ref{avav})
the higher-order terms in Eq.(\ref{hopping2})
enhance the metallic phase, so the MI phase transition line 
is expected to be located in very underdoped region $\delta \ll 1$.

\section{Phase structure of the full model: 
SC transition}
\setcounter{equation}{0}

\begin{figure}[b]
\begin{center}
\includegraphics[width=0.70\hsize]{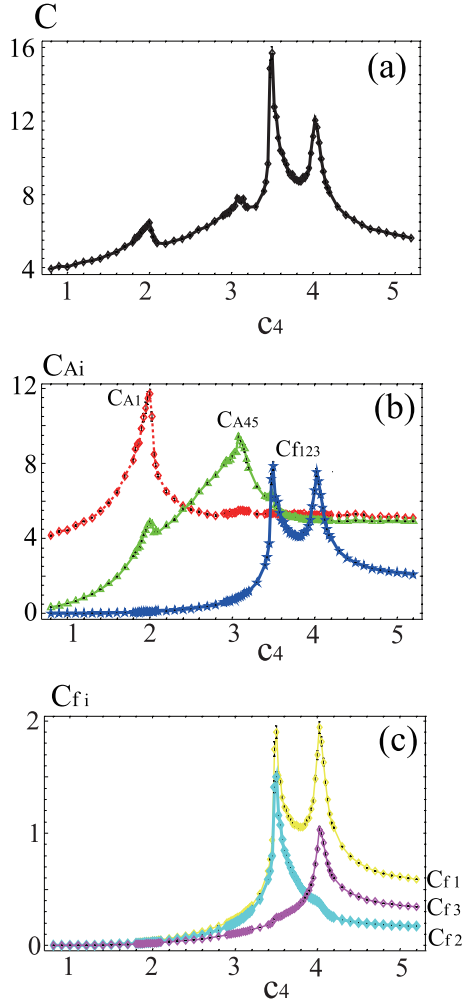}
\caption{Specific heat (a) $C$, (b) $C_{A_i}$, (c) $C_{f_i}$ as functions of 
$c_4$ for $c_1=4.0,\ c_5 = c_4/3.0$ and $L=24$.
$C_{A_{45}}$ implies the fluctuatiuon of $A_4+A_5$, and
$C_{f123}$ for $A_{M}=A_{f_1}+A_{f_2}+A_{f_3}$.
}
\label{fig9}
\end{center}
\end{figure} 

In this section, we study the full model of Eq.(\ref{Afull}) with the 
action $A_{\rm full}=A_{\rm AF}+A_{\rm V}+A_{\rm M}$.
Besides the AF and MI transitions observed in the previous section,
we expect that the new term $A_{\rm M}$ in the action generates 
condensation of the hole-pair field $M_{x\mu}$ 
and/or the holon-pair field $M^\star_{x\mu}$ as the hole density
$\delta$ is increased. This condensation implies generation of a 
SC state.   

We studied the system $A_{\rm full}$ by means of the MC simulations.
As $M^\star_{x\mu}$ is a composite of holons at $x$ and $x+\mu$, 
we put $f_{1,2,3}\propto \delta^2\propto c^2_4$ and 
$M^\star_{x\mu}\in U(1)$ as explained in Sect.II.
Physically, the proportional constants $f_i/c^2_4$ $(i=1,2,3)$ 
depend on the density of holes that actually participate in the SC fluid.
We studied the system $A_{\rm full}$ for various values of $f_i/c^2_4$
and found that the system is stable only for the case with small
values of $f_i/c^2_4$.
For example, the AF phase
disappears at very small value of $c_4$ for $f_i/c^2_4 \sim O(1)$.
In this section, we explicitly show the results for the case 
with $f_1=f_2=f_3=0.03\, c^2_4$.

Let us first study  the high-$T$ region first by choosing 
$c_1=4.0,\ c_5 = c_4/3.0$.
In Figs.\ref{fig9}, we present various specific heat 
as functions of $c_4$.
The total specific heat $C$ in Fig.\ref{fig9}a
exhibits  four peaks at $c_4\simeq 2.0,\ 3.2,\ 3.5$ and $4.1$.
In order to identify the physical meaning of each peak, 
we show the individual specific heat 
$C_{A_i}$ in Fig.\ref{fig9}b and $C_{f_i}\ (i=1,2,3)$ for the
$f_i$-term in $A_M$ defined similarly to Eq.(\ref{partialc})
in Fig.\ref{fig9}c.
From these results, 
it is expected that the first two peaks correspond to the AF 
transition at $c_4\simeq 2.0$  and  the
MI transition at $c_4\simeq 3.2$.
The remaining two terms correspond to 
fluctuations of $f$-terms in the action, and therefore the SC
phase transition.
More precisely, the third peak at $c_4\simeq 3.5$ in $C$ 
corresponds to the $f_1,\ f_2$-terms
and the fourth one at $c_4\simeq 4.1$ to the $f_1,\ f_3$-terms. 
We shall comment on them  later.

\begin{figure}[t]
\begin{center}
\includegraphics[width=0.75\hsize]{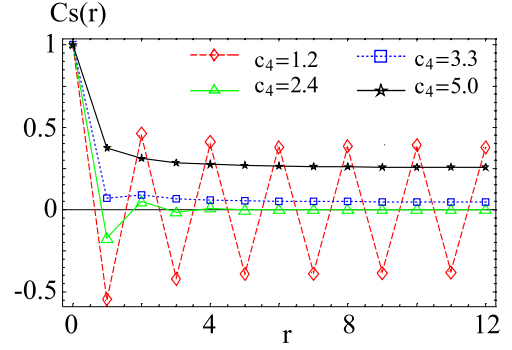}
\caption{Spin correlation functions for various values of $c_4$ for $c_1=4.0,\
c_5 = c_4/3.0$.
$L=24$. At $c_4=1.2$ there is an  AF order; 
At $c_4=2.4$, no magnetic order;
At $c_4=3.3$, a tiny FM order;
At $c_4=5.0$,  a  FM order. }
\label{fig10}
\end{center}
\end{figure}

In Fig.\ref{fig10} we present the spin correlation function
$G_s(r)$ for various values of $c_4$.
It is obvious that only at $c_4=1.2$ the AF LRO exists.
At $c_4=5.0$,  there is a solid  FM order.
This is consistent with the interpretation of four peaks above.

\begin{figure}[t]
\begin{center}
\includegraphics[width=0.7\hsize]{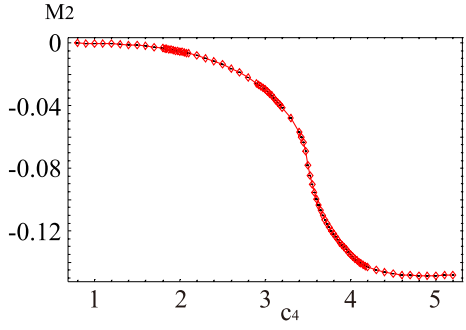}
\caption{Expectation value $M_2$ of Eq.(\ref{M2})
for $c_1=4.0,\ c_5 = c_4/3.0$, $L=24$.
The result shows that $d$-wave correlation between adjacent hole-pair fields
starts to appear at $c_4\sim 3.5$.}
\label{fig11}
\end{center}
\end{figure} 

In order to study the symmetry of SC state,
we consider  the quantity $M_2$, the expectation value
of $M_{x\mu}\bar{M}_{x+\mu,\nu}\ (\mu\neq \nu)$, defined as
\be
M_2\equiv 
\frac{1}{8}\big\la &&M_{x+1,2}\bar{M}_{x1}+\bar{M}_{x+2,1}{M}_{x+1,2}\nn
&&+\bar{M}_{x2}{M}_{x+2,1}+\bar{M}_{x2}M_{x1}\big\ra + {\rm c.c.}
\label{M2}
\ee
In Fig.\ref{fig11} we present $M_2$.
It takes negative values and starts to develop significantly
at $c_1\sim 3.5$, i.e., at the third peak of $C$.
It is obvious that a $d$-wave correlation between adjacent hole-pair fields
is generated beyond the third peak. 

\begin{figure}[b]
\begin{center}
\includegraphics[width=0.8\hsize]{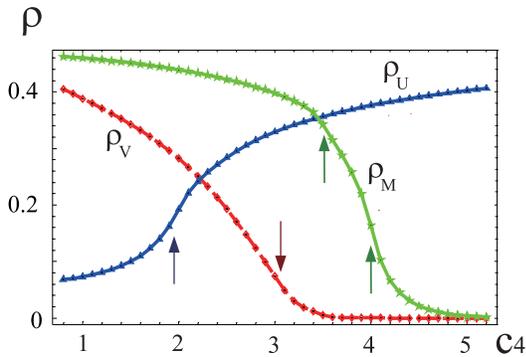}
\caption{Instanton densities $\rho_{U},\ \rho_{V}$
and $\rho_{M^\star}$ as a function of $c_4$ for $c_1=4.0,\
 c_5 = c_4/3.0$ and $L=12$.
Arrows indicate the locations of four peaks in $C$ of
Fig.\ref{fig9}. }
\label{fig12}
\end{center}
\end{figure}

We also measured the instanton densities $\rho_U,\
\rho_V$ and $\rho_{M^\star}$.
$\rho_{M^\star}$ is defined in a similar manner  to $\rho_U$
but by using the
holon-pair field $M_{x\mu}^\star (\equiv M_{x\mu}U_{x\mu}
\sim \bar{\psi}_{x+\mu}\bar{\psi}_x)$ instead of $U_{x\mu}$. 
$\rho_{M^\star}$ reflects
the vortex density of $M_{x\mu}^\star$.
These three instanton densities  are shown in Fig.\ref{fig12}.
By comparing Fig.\ref{fig12} with Fig.\ref{fig4} we see that
the behavior of $\rho_{U}$ and $\rho_{V}$ 
is not influenced strongly by the existence of the $f$-terms,
i.e., $\rho_{U}$ starts to increase at $c_4\simeq 2.0$ and 
$\rho_{V}$ vanishes at $c_4\simeq 3.2$.
The $M^\star$-instanton density $\rho_{M^\star}$ rapidly starts 
to decrease at $c_4\simeq 3.5$
and vanishes at $c_4\simeq 4.1$.
We think that the SC phase transition, which is 
signaled by vanishingly small $\rho_{M^\star}$, 
takes place at $c_4\simeq 4.1$.

\begin{figure}[b]
\begin{center}
\includegraphics[width=0.7\hsize]{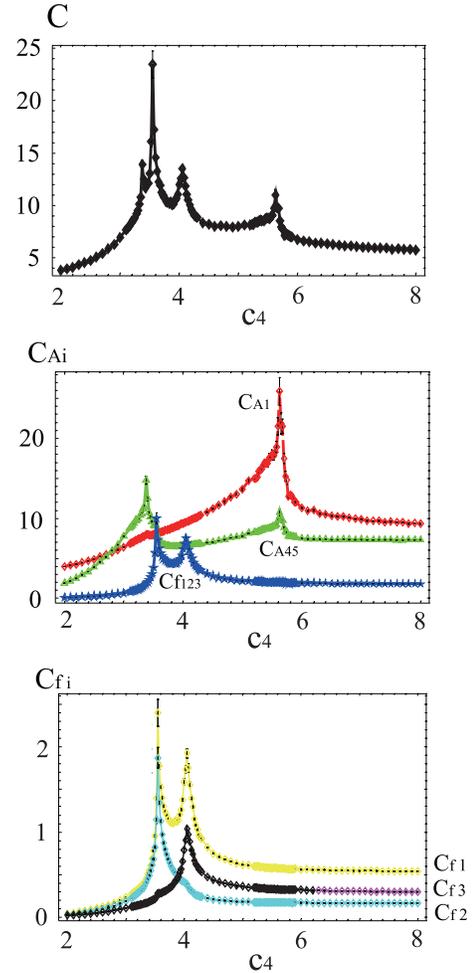}
\caption{Specific heat $C$,\ $C_{A_i}$ and $C_{f_i}$ 
as functions of $c_4$ for $c_1=6.5,\ c_5 = c_4/3.0$
and $L=24$.
}
\label{fig13}
\end{center}
\end{figure} 

From the above numerical calculations, we understand 
the physical meanings of the two peaks  at $c_4 \simeq 3.5,\ 4.1$ 
as follows. 
The third peak at $c_4 \simeq 3.5$ 
are generated by the $f_1$ and $f_2$-terms and is located 
just after the MI transition at $c_4 \simeq 3.2$.
After the MI transition, the holon-hopping amplitude $V_{x\mu}$ becomes
stable, and as a result, these $f_1$ and $f_2$ terms start
to correlate the
phases of {\em a pair of adjacent link fields} $M^\star_{x\mu}$.
In fact,   these two terms in Eq.(\ref{AM}) need a stabilized
$V_{x\mu}$ to let $M_{x\mu}^\star$ stabilize.
Fig.\ref{fig12} shows that
$\rho_{M^\star}$ at $c_4 \simeq 3.5$ is still large.
So this effect is not strong enough to stabilize the holon-pair 
field $M^\star_{x\mu}$ completely at this region of $c_4$.
In order to suppress vortex excitations of the holon-pair field 
$M_{x\mu}^\star$ (making $\rho_{M^\star}$ small enough),
sufficient amount of the $f_3$-term is necessary.
The fourth peak of $C$ at $c_4\simeq 4.1$ 
corresponds to the critical value of $f_3$ 
to realize  such $M^\star_{x\mu}$ stabilization with phase coherence
and generation of SC.
These consideration leads to our conclusion that the genuine SC starts
at the fourth peak $c_4 \simeq 4.1$.

From the above consideration, we expect that the region between the
third and fourth peaks corresponds to a primordial SC state.
In this region we expect that holons acquire a pseudo-gap.
In fact, as $M^\star_{x\mu}$ couples to $\psi_x$ as 
$M^\star_{x\mu}\psi_{x+\mu}\psi_x$, finite expectation value of 
$M^\star_{x\mu}$ supplies fermion-number nonconserving hopping processes
effectively. Together with the fermion-number preserving hopping term
supplied in $A_{{\rm hop}}$ these processes give rise to
a gap in excitation energy of holons\cite{gap}.

Furthermore, from the local gauge symmetry of the system,
the terms like $\bar{M}^\star_{x\mu}z_{x+\mu}z_x$ are
also to be generated by the renormalization effect of high-energy modes
of $z_x$ and $\psi_x$.
Then the spinon field $z_x$ also acquires an extra contribution
to its pseudo-gap, irrespective of a possible pseudo-gap
expected by the mixing of two channels, $c_1U_{x\mu}z^\star_{x+\mu}z_x$
and $c_4V_{x\mu}\bar{z}_{x+\mu}z_x$.
Anyway, the physical properties of that state such as 
excitation spectrum is 
interesting and should be reserved  as a future problem.

\begin{figure}[tb]
\begin{center}
\includegraphics[width=0.7\hsize]{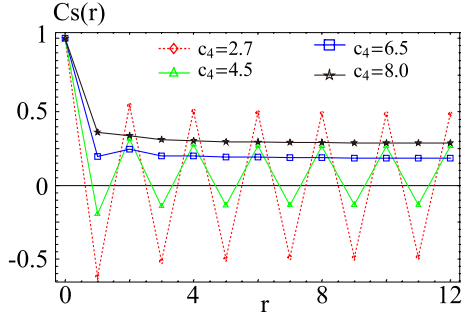}
\caption{Spin correlation functions $G_s(r)$ 
for various values of $c_4$ with $c_1=6.5,
c_5 = c_4/3.0$. 
At $c_4=2.7$ there is an  AF order; 
at $c_4=4.5$, an AF order in a FM background;
at $c_4=6.5$, and $c_4=8.0$, a  FM order. 
}
\label{fig14}
\end{center}
\end{figure} 

\begin{figure}[b]
\begin{center}
\includegraphics[width=0.7\hsize]{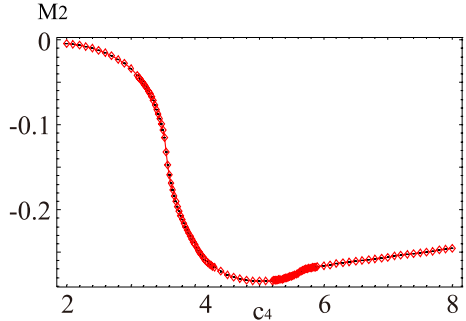}
\caption{Expectation value $M_2$ for $c_1=6.5,\ c_5 = c_4/3.0$, and
$L=24$.
The result shows that $d$-wave correlation between adjacent hole-pair 
fields appears at $c_4\sim 3.6$.
It is interesting to observe that the correlation decreases slightly
in the region {\em without} the AF LRO $c_4>5.6$.}
\label{fig15}
\end{center}
\end{figure} 

\begin{figure}[t]
\begin{center}
\includegraphics[width=0.8\hsize]{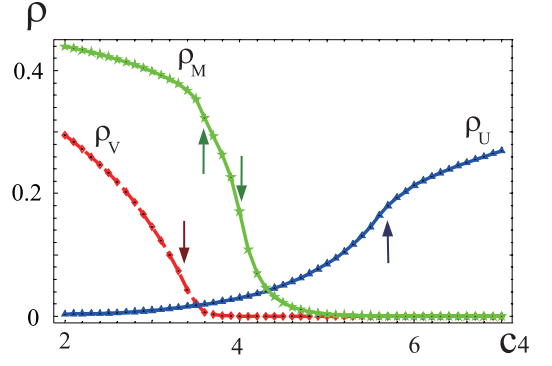}
\caption{Instanton densities $\rho_{U},\ \rho_{V}$
and $\rho_{M^\star}$ as  functions of $c_4$ for $c_1=6.5,\ c_5 = c_4/3.0$ 
and $L=12$. 
Arrows indicate the phase transition points
determined by the specific heat (See Fig.\ref{fig9}).
Their behavior is consistent with the 
phase transition discussed in the text based on Fig.\ref{fig9}.}
\label{fig16}
\end{center}
\end{figure}

Next, let us study the system $A_{\rm full}$ 
at lower-$T$ region by setting $c_1=6.5,\ c_5 = c_4/3.0$.
Behavior of various specific heats, $C,\, C_{A_i},\, C_{f_i}$ 
are shown in Fig.\ref{fig13}.
There are again four peaks in the total specific heat $C$ 
at $c_4\simeq 3.4,\ 3.6, \ 4.0$ and $5.6$.
From the behavior of $C_{A_i}$ and $C_{f_i}$ 
the first peak at $c_4 \simeq 3.4$
corresponds to the MI transition, the peak(s) at $c_4 \simeq 4.0 $(and 3.6) 
to the SC transition,
and the fourth peak at $c_4 \simeq 5.6$ to the AF transition.
The order of these transitions is different from that at the previous 
high-$T$ case as we have already seen in the $UV$ model.
In order to verify the above identification, we calculated the spin 
correlation functions $G_s(r)$, 
the expectation value of adjacent hole-pair field $M_2$,
and the instanton densities  as before.
We show the results in Figs.\ref{fig14}, \ref{fig15} and \ref{fig16}.
These results  support the interpretation of each phase 
given above.

\begin{figure}[b]
\begin{center}
\includegraphics[width=0.7\hsize]{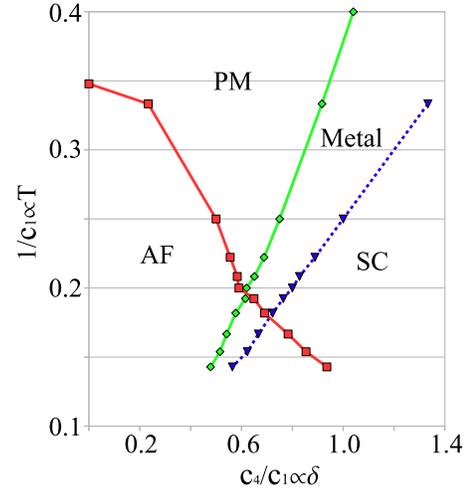}
\caption{Phase diagram of the full model of Eq.(\ref{Afull}) 
in the $\delta$-$T$ plane. The three lines for AF transition,
MI transition and SC transition separate each phase.
All the transitions are of second order.
One may add the pseudo-gap transition line as the fourth line.
}
\label{fig17}
\end{center}
\end{figure} 

In Fig.\ref{fig17}, we present the obtained phase diagram of the
full model $Z_{\rm full}$ of Eq.(\ref{Afull})
in the $\delta-T$ plane.
Each phase are separated by  three transition lines
for AF, MI, and SC transitions.
The SC phase always exists inside the metallic phase, whereas
there is the coexisting phase of the AF and SC at the low-$T$ region.
In addition to these three lines, one may add the line corresponding to the
primordial SC transition as the line of pseudo-gap generation. 
Except for the pseudo-gap transition, which seems 
not to be a sharp transition in experiments,
this phase diagram is consistent with  that observed experimentally
for homogeneous clean underdoped samples\cite{exp2}.

\section{Conclusion and discussion}

In the present paper, we have studied the 
phase structure in the underdoped
region of the $t$-$J$ model by using the slave-fermion representation.
In this formalism, the AF-insulator phase naturally appears
and it is expected that beyond a critical hole concentration $\delta_{\rm MI}$
the coherent hopping of holes is generated and the system
enters into the metallic phase.
This phenomenon was previously studied by the mean-field theory, and
the critical hole concentration was estimated as $\delta_{\rm MI}=0$\cite{su}.

We investigated the system by integrating out the fermionic
holon field by the hopping expansion, which is legitimated for
the region in and near the insulating phase, and then numerically studied 
the AF, MI and SC phase transitions.
The obtained phase diagram is consistent with
that observed experimentally
for clean and homogeneous samples at small hole concentrations.
The present study also implies that the observed pseudo-gap corresponds to a
primordial formation of SC order parameter $M_{x\mu}$.

For the SC phase transition,
we have treated the coefficients of effective action
in more flexible manner than the original hopping expansion
although we maintain the structure of interaction terms.
As explained, this is because these coefficients certainly 
acquire renormalization and even change their signature
as we go into the SC state.
Some of our results in the present paper may reflect this flexibility, 
i.e., they may not be possible in the original $t$-$J$ model
due to the restrictions among the coefficients. 
The  pseudo-gap transition might disappear
(merge to the genuine SC transition) with  different
treatments of the coefficients. 
Even in such case, the results obtained in
the present paper have important meaning as the knowledge of
a reference system to the $t$-$J$ model and other canonical models of the
high-$T_c$ materials.   

Concerning to the SC order parameter, we proposed 
gauge-invariant $M_{x\mu}$ for hole pairs as a most direct 
possibility\cite{im}.
We have calculated its correlation function
$\la\bar{M}_{x\mu}M_{y\nu}\ra$, but
found no LRO of $M_{x\mu}$ even in the SC phase.
We understand this in the following way.
If one could calculate this correlation of the $t$-$J$ model exactly,
one would have LRO in the SC state.
The effective model in  exact treatment certainly contains
a lot of nonlocal interaction terms among $M_{x\mu}$,
although their coefficients are small. Nonvanishing
LRO is to be supported by these nonlocal interactions.
However, the present model truncates the effective interaction terms
to short-range ones, and so fails to produce LRO of $M_{x\mu}$.

However, the study of lattice gauge theory\cite{Kogut} provides us with
a viable alternative of describing  a SC state. 
An effective system may involve only short-range
interactions but it may generates the Higgs phase
in which Meissner effects takes place actually.
The price to pay is that there are no local order parameters
to signal LRO.
Our present model with the action $A_M$ is just a such model.
Because our gauge-noninvariant $M^\star_{x\mu}$ for holon pairs
has vanishing correlations due to gauge-invariant action 
due to Elitzur's theorem\cite{Elitzur}, one need to introduce 
complicated nonlocal order parameters\cite{nonlocal} 
to show that some kind of LRO exists.
Here we note that existence of LRO is a beautiful theoretical
criteria to demonstrate SC phenomenon, but not a necessary condition.
A simple and  direct proof of a SC state may be  to
measure the mass of external  electromagnetic field
and demonstrate the Meissner effect, i.e., the Higgs mechanism.  
We have not made such a proof, but 
the existence of anomalous peak of the specific heat 
certainly demonstrates a new phase, which should corresponds
to the Higgs phase. 
In fact, we have considered a U(1) Ginzburg-Landau model\cite{u1rvb},
which is obtained from $A_M$ of Eq.(\ref{AM}) 
by putting $V_{x\mu}$ to a certain constant.
So the model loses gauge symmetry or viewed as a gauge-fixed version.
The MC simulation of this model certainly exhibits
a Higgs phase for sufficiently large $f_i$ 
in which the correlation functions 
$\la \bar{M}^\star_{x\mu}M^\star_{y\nu}\ra$ exhibit a LRO.
Let us summarize the situation.
Because the faithful effective model of $M_{x\mu}$
is full of nonlocal interactions,
we replace it by a short-range model.
By sacrificing the LRO of gauge-invariant {\it local} order parameter,
we are able to obtain  the new phase. The analysis of the related model
and the experience of lattice gauge theory strongly indicate
this phase is a Higgs phase which is necessary to support SC.

The reason why we integrate out the fermionic holon field analytically 
is obvious,
i.e. it is technically difficult to study fermion systems
by numerical methods.
In recent years, however, it has become possible to numerically simulate
relativistic fermion systems and therefore it is important and also
interesting to study the MI phase transition in the present system
by means of those simulation methods. 
This problem is under study and we hope that the result will be
reported in a future publication.
Even in such a situation, the content of the present paper
may be useful as some basis and a reference to obtain further understanding
of physics of high-$T_c$ superconductors.

\bigskip
\begin{center}
{\bf Acknowledgment} \\
\end{center}
This work was partially supported by Grant-in-Aid
for Scientific Research from Japan Society for the 
Promotion of Science under Grant No.20540264.\\

\appendix

\section{Holon-field integration}
\renewcommand{\theequation}{A.\arabic{equation}} 
\setcounter{equation}{0}

In this appendix, we show some details of holon-field integration
to derive Eq.(\ref{hopping}).
The same techniques are applicable to derive $A_{\rm M}$ in Eq.(\ref{AM}).
It is useful to start with the original path-integral expression\cite{im}
in which Grassmann number $\psi_x(\tau)$ is 
a function of the imaginary-time $\tau$. This is because the ordering
of variables is crucial to obtain the correct results.
Then the relevant integration reads as 
\be
&&\int d\psi_xd\psi_{x+\mu}\exp
\Big[{c_3\over 2\beta}\int_0^\beta d\tau \
(\bar{z}_{x+\mu}z_x)\bar{\psi}_x\psi_{x+\mu}(\tau)
+\mbox{c.c.}\Big] \nn
&&\hspace{0.5cm}=\Big({c_3 \over 2\beta}\Big)^2|\bar{z}_{x+\mu}z_x|^2 \nn
&&\hspace{0.5cm}\times \int_0^\beta d\tau_1d\tau_2 
\langle \bar{\psi}_{x+\mu}(\tau_1)\psi_x(\tau_1)
\bar{\psi}_x(\tau_2)\psi_{x+\mu}(\tau_2)\rangle \nn
&&\hspace{0.5cm}=-\Big({c_3 \over 2\beta}\Big)^2|\bar{z}_{x+\mu}z_x|^2
\int_0^\beta d\tau_1d\tau_2 
\langle \psi_{x+\mu}(\tau_2)\bar{\psi}_{x+\mu}(\tau_1)\rangle \nn
&&\hspace{0.5cm}\times \langle \psi_x(\tau_1)\bar{\psi}_x(\tau_2)\rangle \nn
&&\hspace{0.5cm}=\delta \Big({c_3 \over 2}\Big)^2|\bar{z}_{x+\mu}z_x|^2,
\label{hopint}
\ee
where we have used the following Green function of the hopping expansion,
\be
&&\langle \psi_x(\tau_1)\bar{\psi}_x(\tau_2)\rangle \nn
&=&{e^{-m(\tau_1-\tau_2)} \over 1+e^{-\beta m}} 
[\theta(\tau_1-\tau_2)-e^{-\beta m}\theta(\tau_2-\tau_1)].
\label{Green}
\ee
In Eq.(\ref{Green}), $m$ is the chemical potential and
there holds the relation, 
\be
\delta=\langle \bar{\psi}_x(\tau+0)\psi_x(\tau)\rangle 
={e^{-\beta m} \over 1+e^{-\beta m}}.
\ee




\begin{references}
\bibitem{star}
The variable $z^\star_x$ should not be confused with
 $\bar{z}_x$ or $z^*_x$, 
the complex conjugate of $z_x$. 
In Ref.\cite{im}, the time-reversed spinon field
$\tilde{z}_{x} \equiv i\sigma_2 \bar{z}^{\rm t}_x,\
\tilde{z}_{x1}=\bar{z}_{x2},\ \tilde{z}_{x2}=-\bar{z}_{x1}$
has been introduced. 
$z^\star_x$ here is just its complex conjugate, i.e., 
$z^\star_{x\sigma} =\bar{\tilde{z}}_{x\sigma}$.
In this paper a bar-symbol implies the 
complex conjugate for a complex number and the Grassmann
conjugate for a Grassmann number. 
We do not use asterisks in this paper.
\bibitem{c2}
We use the coefficient $c_3$ instead of $c_2$.
This is partly because
 $c_2$ has been reserved
 for the coefficient of plaquette term
 $\bar{U}_{x\nu}\bar{U}_{x+\nu,\mu}U_{x+\mu,\nu}U_{x\mu}$,
 a possible interaction term that may be 
 included in the generalization of the model.
In this viewpoint, we set $c_2=0$ in this paper.
\bibitem{irrelevant}
Their explicit form and roles are
discussed in details in Ref.\cite{im}.
\bibitem{utrsf}
We should remark that the signatures in front of $\lambda_x$ and
$\lambda_{x+\mu}$ in  the exponent of $U_{x\mu}$
are the same and not the opposite. 
This is in contrast with the usual lattice gauge theory.
\bibitem{FNf3}Strictly speaking, $f_3 \propto \delta^4$
in the leading order.
Here we take the simpler assumption for simplicity of the numerical study.
\bibitem{gap}
The mechanism discussed here to generate a gap 
is a standard one
through a mixing of normal channel and SC channel.
A typical example is the Bogoliubov transformation in 
the Bardeen-Cooper-Schriefer model of
conventional superconductivity. 
For bosonic spinon excitations, one needs 
a  bosonic version of Bogoliubov transformation.
\end{references}
\end{document}